\documentclass[11pt,a4paper,fleqn]{article}
\catcode`\@=11
\@addtoreset{equation}{section}
\def\theequation{\arabic{section}.\arabic{equation}}
\def\appendix{\renewcommand{\thesection}{\Alph{section}}\setcounter{section}{0}
              \renewcommand{\theequation}
            {\mbox{\Alph{section}.\arabic{equation}}}\setcounter{equation}{0}}

\def\maketitle{\thispagestyle{empty}\setcounter{page}0\newpage
                \renewcommand{\thefootnote}{\arabic{footnote}}
                  \setcounter{footnote}0}
\renewcommand{\thanks}[1]{\renewcommand{\thefootnote}{\fnsymbol{footnote}}
               \footnote{#1}\renewcommand{\thefootnote}{\arabic{footnote}}}

\renewcommand{\title}[1]{\begin{center}\Large\bf #1\end{center}\rm\par\bigskip}

\renewcommand{\author}[1]{\begin{center}\Large #1\end{center}}
\newcommand{\address}[1]{\begin{center}\large #1\end{center}}
\newcommand{\pacs}[1]{\smallskip\noindent{\sl PACS numbers:
                       \hspace{0.3cm}#1}\par\bigskip\rm}
\def\babs{\hrule\par\begin{description}\item{Abstract: }\it}
\def\eabs{\par\end{description}\hrule\par\medskip\rm}
\renewcommand{\date}[1]{\par\bigskip\par\sl\hfill #1\par\medskip\par\rm}
\newcommand{\ack}[1]{\par\section*{Acknowledgments} #1}

\newcommand{\dinfn}[1]{${}^{(#1)}$ Dipartimento di Fisica, Universit\`a di Trento\\
                           and Istituto Nazionale di Fisica Nucleare,\\
                                   Gruppo Collegato di Trento, Italia \medskip}

\newcommand{\csic}[1]{${}^{(#1)}$ Consejo Superior de Investigaciones Cient\'\i ficas ICE/CSIC-IEEC,\\
                  Campus UAB, Facultat de Ci\`encies,\\
                      Torre C5-Parell-2a pl, 08193 Bellaterra (Barcelona) Spain\medskip}

\newcommand{\guido}[1]{Guido Cognola$\,{}^{(#1)}$\thanks{e-mail: \sl cognola@science.unitn.it\rm}}
\newcommand{\sergio}[1]{Sergio Zerbini$\,{}^{(#1)}$\thanks{e-mail: \sl zerbini@science.unitn.it\rm}}
\newcommand{\emilio}[1]{Emilio Elizalde$\,{}^{(#1)}$\thanks{e-mail: \sl elizalde@ieec.uab.es\rm}}

\def\segue{\qquad\Longrightarrow\qquad} 
\def\hs{\qquad}               
\def\nn{\nonumber}            
\def\beq{\begin{eqnarray}}    
\def\eeq{\end{eqnarray}}      
\def\ap{\left.}               
\def\at{\left(}               
\def\aq{\left[}               
\def\ag{\left\{}              
\def\cp{\right.}              
\def\ct{\right)}              
\def\cq{\right]}              

\def\R{{\hbox{{\rm I}\kern-.2em\hbox{\rm R}}}}   
\def\H{{\hbox{{\rm I}\kern-.2em\hbox{\rm H}}}}   
\def\N{{\hbox{{\rm I}\kern-.2em\hbox{\rm N}}}}   
\def\C{{\ \hbox{{\rm I}\kern-.6em\hbox{\bf C}}}} 
\def\Z{{\hbox{{\rm Z}\kern-.4em\hbox{\rm Z}}}}   
\def\ii{\infty}                                  
\def\X{\times\,}                                 
\def\dir{/\kern-.7em D\,}                          
\def\lap{\Delta\,}                                 
\def\al{\alpha}\def\ga{\gamma}\def\de{\delta}\def\ep{\varepsilon}
\def\si{\sigma}

\def\Ga{\Gamma}\def\La{\Lambda}

\renewcommand{\title}[1]{\begin{center}\Large\bf #1\end{center}\rm\par\bigskip}
\renewcommand{\author}[1]{\begin{center}\Large #1\end{center}}

\begin{document}

\title{One-loop Euclidean Einstein-Weyl gravity in de Sitter universe}
\author{\guido{a}, \emilio{b}, \sergio{a}}
\address{\dinfn{a} \\ \csic{b}}

\begin{abstract}
By making use of the background field method, the one-loop
quantization for Euclidean Einstein-Weyl  quadratic gravity model  on the de Sitter
universe is investigated. Using generalized zeta function
regularization, the on-shell and off-shell one-loop effective  actions are explicitly
obtained and one-loop renormalizability, as well as the corresponding one-loop renormalization
group equations, are discussed. The so called critical gravity is also considered.

\end{abstract}

\maketitle

\section{Introduction}

Recent astrophysical data indicate that our universe is currently in a
phase of accelerated expansion. The physical origin of this acceleration is not 
completely understood and the related issue is commonly called the dark energy problem.

Several possible explanations have been proposed in the literature. One of them is
based on the use of modified gravitational models,
the simplest one consisting in the inclusion of a small and
positive cosmological constant.
Such model works quite well, according to the most recent data, 
but nevertheless it has some drawbacks
(see, for example \cite{review,rev2,fara10,toni} and reference therein).

Roughly speaking, the idea underlying modified gravity models
is that the Einstein-Hilbert action gives only an approximate
low energy contribution to gravitation and 
additional terms depending on the quadratic curvature invariants should necessarily be included.
This idea is quite old since it was already contained in the seminal paper \cite{staro},
where quadratic terms in the curvature, justified by quantum effects, were added to the
Einstein-Hilbert Lagrangian (for a review, see \cite{buch}).
The important recent finding is that the inclusion of suitable higher order contributions 
may realize not only the actual accelerated
expansion, but also  the early time inflation epoch \cite{seba08}.
Within this context, the de Sitter (dS) space-time plays a fundamental role,
being able to provide an acceleration at different stages in the cosmological set up.

In  previous papers \cite{cogno,guido2,cogno8,cogno9}, $f(R)$ gravity models
and a non local Gauss-Bonnet gravity model at one-loop level in a de Sitter
background have been investigated.
A similar program for the case of  pure Einstein gravity was initiated in
Refs.~\cite{perry,duff80,frad} (see also \cite{ds1,vass92}). Furthermore, such approach
 also suggests  a possible way of understanding the cosmological constant
issue \cite{frad}. Hence, the study of one-loop generalized modified gravity  is a
natural step to be undertaken for the completion of such program,
with the aim to better understand the role and the origin of quadratic corrections in the curvature.
An alternative approach, which is in some sense alternative, has been proposed by Rueter and collaborators \cite{mart}, see also  the review paper 
\cite{martin}, and \cite{martin1}, in which quantum gravity effects in astrophysics and  cosmology are presented.

In the present paper, we will investigate in some detail a model
described by a Lagrangian density which depends on geometric quadratic invariants.
The quantization of quadratic models  of gravity has been discussed in many papers,
and in particular studied in detail on flat space in the seminal paper \cite{stelle}.
A preliminary discussion of a quadratic model based on one-loop on-shell
results has been presented in \cite{cognola12}.

Here we start with the classical Euclidean gravitational action
\beq
I_E[g]=-\int\:d^4x\,\sqrt{g}\,F(R,P,Q)=-\int\:d^4x\,\sqrt{g}\,[M^2R-2\alpha+b(R^2-3P)+\beta G]\,,
\label{action0}\eeq
where $b$ and $\beta$ are dimensionless  parameters,
$M^2$ a mass-squared parameter playing the role of gravitational coupling constant,
and $\alpha$ a ``cosmological constant'' dimensional term.
By $G$ we indicate the Gauss-Bonnet topological invariant which, in 4-dimensions, does
not contribute to the classical field equations.
For this reason, the action in (\ref{action0}) in classically equivalent to the
so called  Einstein-Weyl gravity, since
the  quadratic Weyl invariant $W$ and  the Gauss-Bonnet invariant $G$ are related by
\beq
G-W=\frac{2}{3}(R^2-3P)\,,\hs\hs G=R^2-4P+Q\,,\hs W=\frac13\,R^2-2P+Q\,,
\eeq
$P$ and $Q$ being 
\beq
P=R^{ij}R_{ij}\,,\hs\hs   Q=R^{ijrs}R_{ijrs}\,,\hs\hs i,i,r,s=0,1,2,3.
\eeq
As we already said above, at classical level the Gauss-Bonnet term does not play any role
and could be dropped off but, as we shall see in the following, it do will play an important role
at quantum level.

The action in the form (\ref{action0}) is quite useful in order to discuss
the so called ``critical gravity'', which corresponds to a particular choice of the
$b$ parameter.
An extensive study of 4-dimensional critical gravity,
in the presence of a negative cosmological constant,
has been recently presented in Refs.~\cite{pope0,pope1,pope2,nojiri12},
where additional relevant references can be found.

It should be noted that one-loop Euclidean quantum gravity in a de Sitter background---as 
was fully exploited in \cite{frad} for the case of Einstein's gravity
in the presence of a cosmological term---presents some
peculiar aspects within the background field method.
First, working with the Euclidean version $S(4)$,
one is dealing with a geometric background associated with a
compact manifold without a boundary. This means that the volume is finite and
can be  expressed as a function of the constant Ricci curvature,
which may be chosen as  background field. A
second important remark is that, in order to discuss the one-loop renormalizability
of the model, as well as the related renormalization group equations, one is forced to
work with the off-shell one-loop effective action.
As a consequence, the Landau gauge appears to be the most convenient one. Besides, the usual effective action calculated in this gauge coincides with the Vilkovisky-De Witt effective action (see, for example, \cite{buch}).

Such approach should be compared with the more traditional one,
nicely reviewed in \cite{visser02},
where the Sakharov induced gravity approach \cite{sakharov}
and its modern variants \cite{dima} have been discussed too.
Conformal gravity has been discussed in \cite{ilya}.
The  ghost absence issue for a very general gravitational quadratic model
on Minkowski space-time has been recently investigated in \cite{ana11}. Furthermore, alternative approach is presented in \cite{percacci}.

Regarding to the choice of regularization, since we are dealing with non flat space-time,
it is almost mandatory (or at least very convenient) to make use of a variant of the
generalized zeta-function regularization \cite{dowker,hawking}
(see also \cite{eli94,byts96,kirsten00}),
and the associated heat-kernel techniques \cite{b,s,v}.
In this way, one may evaluate the one-loop
effective action and then study the possibility of stabilization of the
de Sitter background by quantum effects.

The  paper is organized as follows.
Section II contains the evaluation of the quantum fluctuation operators relevant
for the one-loop calculations to be carried out.
In Section III, the off-shell one loop partition function is presented 
and the corresponding one loop renormalization is discussed. Finally, Section IV
is devoted to conclusions.

\section{Quantum field fluctuations around maximally symmetric instantons}

In this Section we will discuss the one-loop quantization of the
model in (\ref{action0}) on a maximally symmetric  space (see, for instance \cite{buch}).
To start with, we consider the Euclidean gravitational action in (\ref{action0})
and, for convenience, we separate linear and quadratic terms
\beq
F(R,P,Q)=f(R)+b(R^2-3P)+\beta G\,,\hs\hs f(R)=M^2R-2\al\,.
\eeq
The model admits a constant Ricci curvature solution $R_0$.
In fact, the general equation for the existence of de Sitter solution \cite{monica,Cognola:2008wy}
\beq
\aq\at\frac12\, R\,\frac{\partial}{\partial R}+P\,\frac{\partial}{\partial P}+
Q\,\frac{\partial}{\partial Q}-1\ct\,F(R,P,Q)\cq_{R=R_0}=0\,,
\label{deSC}\eeq
is trivially satisfied, and reads
\beq
f(R_0)-\frac12\,R_0\,f'(R_0)=0\segue R_0=\frac{4\alpha}{M^2}\,.
\label{AAA1}
\eeq

We are interested in studying quantum fluctuations around the Euclidean dS instanton $S^4$ with positive constant scalar
curvature $R_0$. This is a maximally symmetric space
having covariant conserved curvature tensors. Its
metric may be written in the form
\beq
ds_E^2=d\tau^2(1-H_0^2r^2)+\frac{dr^2}{(1-H_0^2r^2)}+r^2dS_2^2\,,
\eeq
$dS^2$ being the metric of the two-dimensional sphere $S^2$ and $H_0$ the Hubble constant.
The finite volume is given by
\beq
V(S^4)=\frac{384\pi^2}{R_0^2}\,,\hs\hs R_0=12H^2_0\,, \hs\hs G_0=24H^4_0\,,
\eeq
while the Riemann and Ricci tensors are 
\beq
R^{(0)}_{ijrs}=\frac{R_0}{12}\at g^{(0)}_{ir}g^{(0)}_{js}-
g^{(0)}_{is}g^{(0)}_{jr}\ct \:, \hs\hs
R^{(0)}_{ij}=\frac{R_0}{4}\,g^{(0)}_{ij}\,.
\label{AAA2}
\eeq
Now let us consider small fluctuations around the maximally
symmetric instanton. For the sake of completeness,
we consider the general action discussed in \cite{cognola12}, but linear in
$P,Q$. Then, we shall restrict to the action (\ref{action0}), at the end of the
computation. For simplicity, we also put $M^2=1$.
When necessary the right units will be easily recovered by dimensional analysis.

We set
\beq
g_{ij}\longrightarrow g_{ij}+h_{ij}\:,\hs
g^{ij}\longrightarrow g^{ij}-h^{ij}+h^{ik}h^j_k+{\cal O}(h^3)\:,\hs
h=g^{ij}h_{ij}\:,
\eeq
where from now on $g_{ij}\equiv g^{(0)}_{ij}$ is the metric of the
maximally symmetric space and,
as usual, indices are lowered and raised by  means of such metric.

Up to second order in $h_{ij}$, one has
\beq
\sqrt{g}\longrightarrow \sqrt{g}\aq 1+\frac12h+\frac18h^2-\frac14h_{ij}h^{ij}+{\cal O}(h^3)\cq
\eeq
and
\beq
 R &\sim& R_0-\frac{R_0}{4}\,h+\nabla_i\nabla_jh^{ij}-\lap h
\nn \\ && +\frac{R_0}{4}\,h^{jk}h_{jk} -\frac14\,\nabla_ih\nabla^ih
-\frac14\,\nabla_kh_{ij} \nabla^kh^{ij} +\nabla_ih^i_k\nabla_jh^{jk}
-\frac12\,\nabla_jh_{ik}\nabla^ih^{jk} \:,
\eeq
where $\nabla_k$ represents the covariant derivative in the unperturbed metric
$g_{ij}$.
More complicated expressions are obtained for the other invariants $P,Q$,
but for our aims it is not necessary to write them explicitly.

By performing a Taylor expansion of the Lagrangian around de Sitter metric,
up to second order in $h_{ij}$, we get
\beq
I_E[g]\sim-\int\:d^4x\,\sqrt{g}\:
\aq F(R_0,P_0,Q_0)+\frac{hX}2+{\cal L}_2\,\cq\,,
\label{AAA3}
\eeq
where ${\cal L}_2$ represents the second-order contribution and $X$
vanishes when the de Sitter existence condition (\ref{deSC}) is satisfied.
For our particular model, $X=[f(R_0)-(1/2)R_0f'(R_0)]/M^2$.

It is convenient to carry out the standard expansion of the tensor
field $h_{ij}$ in irreducible components \cite{frad}, namely
\beq
h_{ij}=\hat
h_{ij}+\nabla_i\xi_j+\nabla_j\xi_i+\nabla_i\nabla_j\sigma
+\frac14\,g_{ij}(h-\lap\sigma)\:,
\label{tt}\eeq
where $\si$ is
the scalar component, while $\xi_i$ and $\hat h_{ij}$ are the vector
and tensor components, with the following properties
\beq
\nabla_i\xi^i=0\:,\hs\hs \nabla_i\hat h^{ij}=0\:,\hs\hs \hat h_i^i=0\:.
\label{AAA4} \eeq
In terms of the irreducible components of the $h_{ij}$ field,
the Lagrangian density, disregarding total derivatives, becomes
\begin{eqnarray}
{\cal L}_2&=&{\cal L}_{hh}+2\,{\cal L}_{h\si}+{\cal L}_{\si\si}+{\cal L}_V+{\cal L}_T\,,
\end{eqnarray}
where ${\cal L}_{hh},{\cal L}_{h\si},{\cal L}_{\si\si}$ represent the scalar
contribution (a $2\X2$ matrix),
while ${\cal L}_V$ and ${\cal L}_T$ represent the vector and tensor contributions, respectively.
One has
\begin{eqnarray}
{\cal L}_{hh}&=&h\,\aq
\frac{1}{32}{F_{RR}}R_0^2
-\frac{1}{32}{F_{R}}R_0
+\frac{X}{16}
-\frac{3}{32}{F_{R}}\Delta
+\frac{1}{16}{F_{P}}R_0\Delta
+\frac{1}{16}{F_{Q}}R_0\Delta
\cp\nonumber\\ &&\ap\hs\hs
+\frac{3}{16}{F_{RR}}R_0\Delta
+\frac{3}{16}{F_{P}}\Delta^2
+\frac{3}{16}{F_{Q}}\Delta^2
+\frac{9}{32}{F_{RR}}\Delta^2
\cq\,h\:,
\end{eqnarray}
\begin{eqnarray}
{\cal L}_{h\si}&=&h\,\aq
-\frac{1}{16}{F_{RR}}R_0^2\Delta
+\frac{1}{16}{F_{R}}R_0\Delta
-\frac{1}{8}{F_{P}}R_0\Delta^2
\cp\nonumber\\ &&\hs\hs
-\frac{1}{8}{F_{Q}}R_0\Delta^2
-\frac{3}{8}{F_{RR}}R_0\Delta^2
+\frac{3}{16}{F_{R}}\Delta^2
\nonumber\\ &&\ap\hs\hs\hs\hs
-\frac{3}{8}{F_{P}}\Delta^3
-\frac{3}{8}{F_{Q}}\Delta^3
-\frac{9}{16}{F_{RR}}\Delta^3
\cq\,\sigma\,,
\end{eqnarray}
\begin{eqnarray}
{\cal L}_{\si\si}&=&\sigma\,\aq
\frac{1}{32}{F_{RR}}R_0^2\Delta^2
-\frac{1}{16} XR_0\Delta
-\frac{1}{32}{F_{R}}R_0\Delta^2
-\frac{3}{16} X\Delta^2
\cp\nonumber\\ &&\hs\hs
+\frac{1}{16}{F_{P}}R_0\Delta^3
+\frac{1}{16}{F_{Q}}R_0\Delta^3
+\frac{3}{16}{F_{RR}}R_0\Delta^3
-\frac{3}{32}{F_{R}}\Delta^3
\nonumber\\ &&\ap\hs\hs\hs\hs
+\frac{3}{16}{F_{P}}\Delta^4
+\frac{3}{16}{F_{Q}}\Delta^4
+\frac{9}{32}{F_{RR}}\Delta^4
\cq\,\sigma\,,
\end{eqnarray}
\begin{eqnarray}
{\cal L}_V&=&\xi^k\,\aq
\frac{1}{8}R_0X+\frac{1}{2}X\Delta
\cq\,\xi_k\:,
\end{eqnarray}
\begin{eqnarray}
{\cal L}_T&=&h^{ij}\,\aq
-\frac{1}{72}{F_{P}}R_0^2
+\frac{1}{36}{F_{Q}}R_0^2
-\frac{1}{24}{F_{R}}R_0
-\frac{1}{4}X
+\frac{1}{4}{F_{R}}\Delta
\cp\nonumber\\ &&\ap\hs\hs
+\frac{1}{24}{F_{P}}R_0\Delta
-\frac{1}{3}{F_{Q}}R_0\Delta
+\frac{1}{4}{F_{P}}\Delta
+{F_{Q}}\Delta^2
\cq\,\hat h_{ij}\,.
\end{eqnarray}
where $\lap=g^{ij}\nabla_i\nabla_j$ is the Laplace-Beltrami operator in the unperturbed
metric $g_{ij}$, which is
a solution of the field equations, but only if $X=0$.
We have written the above expansions around a maximally symmetric space,
which in principle would not be a solution. This means, in other words, that the function
$f(R)$ can be arbitrary. In the latter expression
$F_R,F_{RR}$ represent the first and second derivatives of $F(R,P,Q)$ with respect to $R$
evaluated on de Sitter metric $g_{ij}$. And similarly for $F_P,F_Q$.

As is well known, invariance under diffeomorphisms renders the
operator in the $(h,\si)$ sector not invertible. One needs a gauge
fixing term and a corresponding ghost compensating term. Here we choose
the harmonic gauge, that is
\beq
\chi_j=-\nabla_ih^i_j-\frac12\,\nabla_j h=0\,,
\eeq
and the gauge fixing term
\beq
{\cal L}_{gf}=\frac12\,\chi^iG_{ij}\chi^j\,,\hs\hs G_{ij}=\ga\,g_{ij}\,.
\label{AAA5}
\eeq
The corresponding ghost Lagrangian reads \cite{buch}
\beq
{\cal L}_{gh}= B^i\,G_{ik}\frac{\de\,\chi^k}{\de\,\ep^j}C^j\,,
\label{AAA6} \eeq
where $C_k$ and $B_k$ are the ghost and anti-ghost
vector fields respectively, while $\de\,\chi^k$ is the variation of
the gauge condition due to an infinitesimal gauge transformation of
the field. In this case, it reads
 \beq
\de\,h_{ij}=\nabla_i\ep_j+\nabla_j\ep_i\segue
\frac{\de\,\chi^i}{\de\,\ep^j}=g_{ij}\,\lap+R_{ij}\,.
\label{AAA7} \eeq
Neglecting total derivatives, one has
\beq {\cal L}_{gh}=B^k\,\ga\,\at\lap+\frac{R_0}{4}\ct\,C_k\,.
\label{AAA8} \eeq
In irreducible components one finally obtains
\begin{eqnarray}
{\cal L}_{gf} &=&\frac{\ga}2\aq\xi^k\,\at\lap+\frac{R_0}4\ct^2\,\xi_k
    +\frac{3\rho}{8}\,h\,\at\lap+\frac{R_0}3\ct\,\lap\,\si
\cp\nn\\&&\hs\ap
    -\frac{\rho^2}{16}\,h\,\lap\,h
-\frac{9}{16}\,\si\,\at\lap+\frac{R_0}3\ct^2\,\lap\,\si\cq
\label{AAA10} \eeq
\beq {\cal L}_{gh} &=&
\ga\aq\hat B^k\at\lap+\frac{R_0}{4}\ct\hat C_k
+\frac{\rho-3}{2}\,\hat b\,\at\lap-\frac{R_0}{\rho-3}\ct\,\lap\hat c\cq\,,
\eeq
where ghost irreducible components are defined by
\beq
C_k&=&\hat C_k+\nabla_k\hat c\,,\hs\hs \nabla_k\hat C^k=0\,,
\nn\\
B_k&=&\hat B_k+\nabla_k\hat b\,,\hs\hs \nabla_k\hat B^k=0\,.
\label{AAA11} \eeq

\section{Off-shell one-loop effective action}

In order to compute the one-loop contributions to
the effective action one has to consider the path integral for the
bilinear part,
${\cal L}= {\cal L}_2+\,{\cal L}_{gf}+{\cal L}_{gh}, $
of the total Lagrangian and take into
account the Jacobian due to the change of variables with respect to
the original ones. In this way, one gets  \cite{frad,buch}
\beq
Z^{(1)}&=&\at\det G_{ij}\ct^{-1/2}\,\int\,D[h_{ij}]D[C_k]D[B^k]\:
\exp\,\at -\int\,d^4x\,\sqrt{g}\,{\cal L}\ct
\nn\\
&=&\at\det G_{ij}\ct^{-1/2}\,\det J_1^{-1}\,\det J_2^{1/2}\,
\nn\\&&\times \int\,D[h]D[\hat h_{ij}]D[\xi^j]D[\si]
D[\hat C_k]D[\hat B^k]D[c]D[b]\:\exp\, \at-\int\,d^4x\,\sqrt{g}\,{\cal L}\ct\,,
\eeq where $J_1$  and $J_2$ are the Jacobians coming from the
change of variables in the ghost and tensor sectors, respectively
\cite{frad}. They read
\beq J_1=\lap_0\,,\hs\hs J_2=\at\lap_1+\frac{R_0}{4}\ct\at\lap_0+\frac{R_0}{3}\ct\,\lap_0\,,
\label{AAA13} \eeq
and the determinant of the operator $G_{ij}$ is trivial in this case.
Here and in the following $\lap_0,\lap_1,\lap_2,$ represent the Laplacian
acting on scalars, vectors and tensors, respectively.

Due to the presence of curvature, the Euclidean gravitational action
is not bounded from below, because  arbitrary negative contributions
can be induced on $R$ by conformal
rescaling of the metric. For this reason
we have also used the Hawking prescription of integrating over
imaginary scalar fields. Furthermore, the problem of the presence of
additional zero modes introduced by the decomposition (\ref{tt}) can
be treated by making use of the method presented in Ref.~\cite{frad}.

Now, for the action (\ref{action0}) a straightforward computation
leads to the off-shell
one-loop contribution to the ``partition function''. In the Landau gauge,
$\rho=1,\ga\to\ii$,  with $X=R_0/2-2\alpha/M^2$, we get
\beq
\Ga_{off-shell}=I_E(g)+\Ga^{(1)}_{off-shell}\,,\hs\hs
                 I_E(g)=96\pi^2\at\frac{2M^2}{R_0}+ b\ct+64\pi^2\beta\,,
\eeq
\beq
\Ga^{(1)}_{off-shell}&=&\sum_i \frac12\,\log\det \frac{L_i}{\mu^2} =
 \frac12\,\log\det\at\frac1{\mu^2}\,\aq-\lap_0-\frac{2\alpha}{M^2} \cq\ct
\nn\\&&\hs
  -\frac12\,\log\det\at\frac1{\mu^2}\,\aq-\lap_1-\frac{R_0}4\cq\ct-\frac12\,\log\det\at\frac1{\mu^2}\,
\aq-\lap_0-\frac{R_0}{2}\cq\ct
\nn\\&&
  +\frac12\,\log\det\at\frac1{\mu^2}\,\aq-\lap_2-Y_{+}\cq\ct
    +\frac12\,\log\det\at\frac1{\mu^2}\,\aq-\lap_2-Y_{-}\cq\ct\,,
\label{offEA}\eeq
where
\beq
Y_{\pm}=\frac{1}{12}\at -3R_0-\frac{2M^2}{b}\pm\frac{1}{b} \sqrt{96b \alpha+4M^4-20bM^2R_0+b^2R_0^2} \ct\,.
\eeq
As usual, an arbitrary renormalization parameter $1/\mu^2$ has
been introduced for dimensional reasons.
As expected, the parameter $\beta$ does not appear in the latter expression, since
the Gauss-Bonnet invariant does not give contributions to the field equations,
but it gives a constant contribution to the classical action,
which will actually play an important role in the renormalization procedure.

\subsection{On-shell one-loop effective action}
As is well known, the on-shell effective action does not have
to  depend on the gauge and, in fact, setting $X=0$, that is
$M^2R_0-4\alpha=0$, we get
\beq
\Ga_{on-shell}&=& 96\pi^2\at\frac{2M^2}{R_0}+ b\ct+64\pi^2\beta
  -\frac12\,\log\det\at\frac1{\mu^2}\,\aq-\lap_1-\frac{R_0}4\cq\ct
\nn \\ 
  &+&\frac12\,\log\det\at\frac1{\mu^2}\,\aq-\lap_2+\frac{R_0}6\cq\ct
    +\frac12\,\log\det\at\frac1{\mu^2}\,\aq-\lap_2+\frac{R_0}{3}+\frac{M^2}{3b}\cq\ct\,.
\label{1-loopEA}\eeq
The above expression is only formal, and one needs regularization. For the moment,
let us imagine to be dealing with the finite part of such an effective action.

We observe that there exists a ``critical'' value for $b$ for which
all spin excitations become ``massless''. In fact, choosing
\beq
b=b_{crit}=-\frac{2M^2}{R_0}=-\frac{M^4}{2\alpha}\,,
\eeq
the effective action simplifies to
\beq
\Ga_{crit}&=&64\pi^2\beta
      -\frac12\,\log\det\at\frac1{\mu^2}\,\aq-\lap_1-\frac{R_0}4\cq\ct
       +\log\det\at\frac1{\mu^2}\,\aq-\lap_2+\frac{R_0}{6}\cq\ct\,.
\label{1-loopEAc}
\eeq
In contrast to the AdS case, in Euclidean dS space $R_0>0$, and so $b_{crit}<0$.

The stability of the dS solution can be investigated by
looking at the spectra of the Laplace-type operators and it then follows that
all eingenvalues are non-negative as in general relativity,
with the possible presence of a zero mode \cite{frad,guido2}.
As a consequence, dS background space is stable,
in agreement with the classical analysis presented in \cite{monica,Cognola:2008wy}.

But what about the stability of the critical values with respect to renormalization?
According to the background field method one should work at the off-shell level.
Nevertheless,  one may try an on-shell, one-loop renormalization,
by observing that $\beta$ might become
a ``bare'' constant, and its redefinition may contain all counterterms
necessary in order to cancel the on-shell one-loop
divergences coming from the functional determinants.
In general, using a variant of the zeta function regularization procedure
\cite{cognola12},  at one-loop level one has
\beq
\Gamma(\mu,\varepsilon)=I_E(\mu,\epsilon)-\frac{1}{2}\sum_i\aq \frac{\zeta(0|L_i)}{\varepsilon}+\zeta(0|L_i)\log \mu^2+
\zeta'(0|L_i)\cq\,,
\label{GammaEp}\eeq
where the summation is over all Laplace-type operators appearing in the one-loop contribution to
the action.

For the critical gravity in (\ref{1-loopEAc}), we have
$L_1=-\lap_1-R_0/4$ and $L_2=-\lap_2+R_0/6$, thus
\beq
\Gamma(\mu,\varepsilon)&=&64\pi^2\at\frac{\beta_0}{\varepsilon}+\beta(\mu)\ct
       + \frac{1}{2\varepsilon}\aq\zeta(0|L_1)-2\zeta(0|L_2)\cq
\nn \\ &&\hs\hs
           +\log\mu\,\aq\zeta(0|L_1)-2\zeta(0|L_2)\cq
             +\frac{1}{2}\,\zeta'(0|L_1)-\zeta'(0|L_2)\,,
\eeq
where $\beta_0$ is the bare coupling constant and $\beta(\mu)$ the running one.
Making a suitable choice for $\beta_0$, one has the renormalized on-shell effective critical action
\beq
\Gamma_{crit}(\mu)=64\pi^2\beta(\mu)+\log\mu\,\aq\zeta(0|L_1)-2\zeta(0|L_2)\cq
               +\frac{1}{2}\zeta'(0|L_1)-\zeta'(0|L_2)\,.
\eeq
The eigenvalues of the Laplace type operators on $SO(4)$ are well known and in this way it is possible to compute the zeta-functions appearing in the expression above explicitly.
In particular,
$\zeta(0|L_1)=-191/30$ and $\zeta(0|L_2)=89/9$, while
$\zeta'(0|L_1)$ and $\zeta'(0|L_2)$ are computable expressions independent on $\mu$.
The usual imposition
\beq
\mu\,\frac{d\Gamma(\mu)}{d\mu}=0\,,
\label{RGE}\eeq
gives rise to the renormalization group equation for critical gravity, in the form
\beq
\mu\,\frac{d\beta(\mu)}{d\mu}=2\zeta(0|L_2)-\zeta(0|L_1)\sim 26>0\,.
\eeq
This is the only running coupling constant and, thus, on-shell critical gravity
seems to be stable at the one-loop level.
But this is not really conclusive since, strictly, the issue of criticality
 depends on the on-shell expression.


\subsection{Off-shell one-loop renormalization}

As far as the  off-shell one-loop renormalization is concerned,
the situation is completely different with respect to the previous one
and apparently there is no room for the notion of criticality.

Again, the starting point is the equation in (\ref{GammaEp}), but now
the classical action contains all the bare quantities,
which generate the counterterms for absorbing the one-loop divergencies. It reads
\beq
I_E(\mu,\epsilon)&=&384\pi^2\aq \frac{M^2(\mu,\varepsilon)}{R_0}
                       -\frac{2\alpha(\mu,\varepsilon)}{R^2_0}
                        +\frac{b(\mu, \varepsilon)}{4}+\frac{\beta(\mu,\varepsilon)}{6}\cq
\nn\\ &=& 384\pi^2\aq\frac{M^2(\mu)}{R_0}-\frac{2\alpha(\mu)}{R^2_0}
            +\frac{b(\mu)}{4}+\frac{\beta(\mu)}{6}\cq
              +\frac{1}{\varepsilon}\aq  \frac{A_1}{R_0}+\frac{B_1}{R^2_0}+ C_1\cq\,,
\eeq
where we have separated the finite and divergent parts of the coupling constants by
means of suitable finite quantities $A_1,B_1,C_1$ independent of $R_0$.
On the other hand, a direct computation shows that
\beq
\sum_i \zeta(0|L_i)=\aq\frac{A(\mu)}{R_0}+\frac{B(\mu)}{R^2_0}+C(\mu)\cq \,,
\label{z0}
\eeq
where $L_i$ are all Laplace-type operators in (\ref{offEA}) and
$A(\mu),B(\mu),C(\mu)$ are finite functions depending on the renormalized running
coupling constants $M^2(\mu),\alpha(\mu),b(\mu)$. They read
\beq
A(\mu)=\frac{8\alpha(\mu)}{M^2(\mu)}\,,\hs B(\mu)=\frac{20M^4(\mu)}{3b^2(\mu)}
+\frac{80\alpha(\mu)}{b(\mu)}+\frac{48\alpha^2(\mu)}{M^4(\mu)}\,,\hs C(\mu)=\frac{1763}{90}\,.
\eeq
The model is one-loop renormalizable since all one-loop divergences can actually be absorbed by
an appropriate choice of  $A_1,B_1,C_1$.
The finite, renormalized one-loop effective action  reads
\beq
\Gamma(\mu)=384\pi^2 \aq \frac{M^2(\mu)}{R_0}-\frac{2\alpha(\mu)}{R^2_0}+\frac{b(\mu)}{4}\cq
    +\log\mu\,\aq\frac{A(\mu)}{R_0}+\frac{B(\mu)}{R^2_0}+ C(\mu)\cq+Z\,,
\eeq
where we have dropped the parameter $\beta(\mu)$ because  here it does not play any role,
and we have set
$Z=-\frac{1}{2}\sum_i\zeta'(0|L_i)$. This is the finite part
of the functional determinant which does not depend on $\mu$ and
in principle can be explicitly evaluated.


As above, the one-loop renormalization group equations can be obtained by
means of (\ref{RGE}). To this aim it is convenient to introduce
the dimensionless variable $\rho=\log\frac{\mu}{\mu_0}$,
$\mu_0$ being a reference low energy scale.
>From (\ref{RGE}) we obtain the three differential equations
\beq\ag\begin{array}{l}
\frac{db}{d\rho}=c_b\,, \\
\frac{d M^2}{d\rho}=\frac{\alpha}{48\pi^2M^2}\,,\\
\frac{d\alpha}{d\rho}=-\frac1{192\pi^2b^2}\,
        \at12\frac{\alpha^2b^2}{M^4}+20\alpha b+\frac53\,M^4\ct\,,
\end{array}\cp\hs\hs  c_b=\frac{1763}{8640\pi^2}\,,
\label{m}\eeq
where all coupling constants are functions of $\rho$.
Solving the system of differential equations above, we finally get
\beq\ag\begin{array}{l}
   b(\rho)=c_b\,\rho+c_0\,,\\
     M^2(\rho)=c_1(\rho+c_0)^{p_1-p_2}\aq(\rho+c_0)^{10p_2}+c_2\cq^{1/5}\,,\\
       \alpha(\rho)=48\pi^2\frac{M^4(\rho)}{(\rho+c_0)}
              \aq p_1-p_2+\frac{2p_2}{1+c_2(\rho+c_0)^{-10p_2}}\cq\,,
\end{array}\cp\hs\hs
\ag\begin{array}{l}
    p_1=\frac{863}{17630}\,,\\ \\
     p_2=\frac{\sqrt{474769}}{17630}\,.
\end{array}\cp
\label{sm}\eeq
The integration constants $c_0,c_1,c_2$ depend on the initial conditions and we assume all of them
to be non negative. Moreover, to simplify the discussion, from now on we shall take $c_2=0$. 
With this assumption we get
\beq
M^2(\rho)=c_1(\rho+c_0)^p\,,\hs\hs
\alpha(\rho)=48\pi^2\,p\,\frac{M^4(\rho)}{\rho+c_0}\,,\hs\hs p=p_1+p_2\sim0.09\,,
\label{sm1}\eeq
and the one-loop, running, gravitational coupling constant reads
\beq
G(\rho)=\frac{1}{16\pi\,M^2(\rho)}=\frac{1}{16\pi\,c_1(\rho+c_0)^p}\,,
\eeq
while the one-loop, running, cosmological constant is
\beq
\Lambda(\rho)=\frac{\alpha(\rho)}{M^2(\rho)}=48\pi^2\,p\,c_1\,(\rho+c_0)^{p-1}\,.
\eeq
As a result,  there is no Landau pole, and at large energy $\mu\gg\mu_0$ or equivalently $\rho\rightarrow\infty$, 
one has that both $G(\rho)$ and $\Lambda(\rho)$ go to zero.  This property is the analogue on de Sitter space of the well known  gravitational 
asymptotic freedom for quadratic gravity \cite{Fra,Tombo,avra,avra1,nara}. 

Furthermore,  we may assume general relativity to be valid at 
low energy, that is
$\mu\sim\mu_0$ or equivalently $\rho\sim0$, then 
\beq
\left.G(\rho)\right|_{\rho\to0}=G_N\segue  
   c_1=\frac{c_0^{-p}}{16\pi\,G_N}\,,\hs 
    M^2(\rho)=\frac{1}{16\pi\,G_N}\at1+\frac\rho{c_0}\ct^p\,,
\eeq
$G_N=G(0)$ being the Newton constant.

To conclude this section, we write down the effective field equation given by
\beq
\frac{\partial \Gamma}{\partial R_0}=0\,.
\eeq
The solution can be written in the implicit form
\beq
R_0=\frac{1}{1-\frac{A\,\rho}{384\pi^2M^2}}
  \aq\frac{4\alpha}{M^2}-\frac{B\,\rho}{192\pi^2M^2}+\frac{R_0^3}{384 \pi^2M^2}\,
      \frac{\partial Z}{\partial R_0} \cq \,.
\eeq
This is a quite complicated expression in the unknown variable $R_0$. 
Of course, at low energy $\rho\sim0$, $R_0\ll M^2$, 
one gets the classical solution 
\beq
R_0 \sim \frac{4\al(0)}{M^2_P(0)}=4\La(0)\,,
\eeq
but in principle other regimes can be studied. 

\section{Conclusions}

In this paper, the Einstein gravity plus a quadratic gravitational Weyl 
term has been investigated by computing the corresponding 
one-loop quantum corrections  by means of the
background field method. 
As a classical background we have considered the de Sitter one 
for its potentially very important physical applications.
The one-loop calculation has been performed in the Euclidean sector,
where the classical background is the compact manifold $S^4$. 
In the calculation, due to the fact that we are working with a non flat background, 
we are forced to make use of (a variant of) zeta-function regularization. 
In the presence of this compact curved manifold, the one-loop effective action, 
and also the ensuing one-loop renormalization group equations, 
have been  computed and carefully investigated. 
On shell, the associated quadratic critical gravity has been discussed. 
In order to investigate the role of the one-loop corrections, we have to consider 
the off-shell one-loop effective action and so,
to get rid of the gauge dependence, the Landau gauge has been used. In this way 
the critical conditions are lost, in general. 
In fact, also in the simplified case we have considered ($c_2=0$), 
the critical ratio $\frac{M^4(\rho)}{2\alpha(\rho)}\sim(\rho+c_0)$,
which is not equal to $b(\rho)$. 
This means that the critical gravity condition are not stable under the one-loop renormalization flow.
As a consequence,   one might doubt about its relevance at 
least when the background is the compact manifold $S^4$. In fact,  in the anti de Sitter case (AdS) the situation could be completely different, 
because the Euclidean counterpart of AdS is the non compact hyperbolic manifold $H^4$. 

In conclusion, it should also be interesting to repeat this one-loop 
calculation in alternative extended gravity models, 
for example, in the so-called $f(T)$ gravity models, which depend on geometric invariants
built up by using torsion field
(for details see \cite{t} and the references therein).

\ack{We would like to thank L.~Vanzo and S.D.~Odintsov for valuable suggestions. Part of this work  was done at the Kobajashi-Maskawa Institute in Nagoya, and S.Z. and E.E. would like to thank all the members of Center for Theoretical Studies for very kind hospitality}. E.E. was partly supported by MICINN (Spain), grant PR2011-0128 and project FIS2010-15640, by the CPAN Consolider Ingenio Project, and by AGAUR (Generalitat de Ca\-ta\-lu\-nya), contract 2009SGR-994.

\end{document}